\documentstyle[12pt,aaspp4,psfig]{article}

\begin{document}

\title{The WOMBAT Challenge:  A ``Hounds and Hares'' Exercise for Cosmology}

\author{
Eric Gawiser\footnote{gawiser@astron.berkeley.edu}$^,$\footnote{Physics Department}, 
Douglas Finkbeiner$^2$, 
Andrew Jaffe\footnote{Center for Particle Astrophysics}, 
Joanne C. Baker\footnote{Astronomy Department}, 
Amedeo Balbi$^{3,}$\footnote{Lawrence Berkeley National Laboratory},
Marc Davis$^{2,3,4}$,
Shaul Hanany$^3$, 
William Holzapfel$^{2,3}$,
Leonidas Moustakas$^4$,
James Robinson$^4$,
Evan Scannapieco$^2$,
George F. Smoot$^{2,3,5}$
\& Joseph Silk$^{2,3,4}$ 
}
\affil{University of California, Berkeley, CA 94720}

%
%
%
%
%
%
\begin{abstract}

The Wavelength-Oriented Microwave Background Analysis Team (WOMBAT) 
 is constructing microwave skymaps 
which will be more realistic than previous simulations.  
Our foreground models represent a considerable
 improvement:    
where spatial templates are available for a given foreground, we 
predict the flux and spectral index of 
that component at each place on the sky and estimate the 
uncertainties in these quantities.  
We will produce maps containing
 simulated Cosmic Microwave Background anisotropies 
 combined with all major expected foreground components.  
  The simulated maps will be provided to the cosmology community 
as the WOMBAT Challenge,  
a ``hounds and hares'' exercise where such maps can be 
analyzed to extract cosmological parameters by scientists who 
are unaware of their input values. 
This exercise will 
test the efficacy of current foreground subtraction, power 
spectrum analysis, and parameter estimation techniques and will help 
identify 
the areas most in need of progress.  

\end{abstract}


\section{Introduction}

Cosmic Microwave Background (CMB) anisotropy observations during the 
next decade will yield data of unprecedented quality and quantity.  
Determination of cosmological parameters to the 
precision that has been forecast (Jungman 
et al. 1996, Bond, Efstathiou, \& Tegmark 1997,
Zaldarriaga, Spergel, \& Seljak 1997, Eisenstein, Hu, \& Tegmark 1998) 
will require significant 
advances in analysis techniques to handle the 
large volume of data, subtract foreground contamination,  
and account for instrumental systematics.  To guarantee accuracy 
we must ensure that these analysis techniques 
do not introduce unknown biases into the estimation of cosmological 
parameters.  

The Wavelength-Oriented Microwave Background Analysis Team 
(WOMBAT, http://astro.berkeley.edu/wombat)
 will produce state-of-the-art 
simulations of microwave 
foregrounds, using all available information about the 
frequency dependence, power spectrum, and spatial distribution of 
each component.  Using the phase information (detailed spatial morphology 
as opposed to just the power spectrum) 
of each foreground component 
 offers the possibility of improving upon foreground 
subtraction techniques that only use the predicted angular 
power spectrum of the foregrounds to account for their spatial distribution.
Most foreground separation techniques rely on assuming that the frequency 
spectra of the components is constant across the sky, but we will provide
 information on the spatial variation of each component's spectral index 
whenever possible.  
The most obvious advantage of this approach 
 is that it reflects our actual sky.  With 
the high precision expected from future CMB maps we must test our 
foreground subtraction techniques on as realistic a sky map as 
possible.  A second advantage is the construction
of a common, comprehensive database for all known CMB foregrounds.  
The database will include 
known uncertainties in the estimation of the foregrounds.  Such a data 
base should prove valuable for all groups involved in measuring 
the CMB and extracting cosmological information from it.  Section 2  
describes our plans to generate foreground models which include 
phase information, 
and Section 3 gives a brief survey of existing subtraction techniques 
and their limitations. 

These microwave foreground models 
provide the perfect starting point for 
the WOMBAT Challenge, 
a ``hounds and hares'' exercise in which we will generate skymaps for 
various cosmological models and offer them to the cosmology community for 
analysis without revealing the input parameters.  
This challenge is similar to the ``Mystery CMB Sky Map challenge'' posted 
by our sister collaboration, 
COMBAT\footnote{Cosmic Microwave Background Analysis Tools, 
http://cfpa.berkeley.edu/group/cmbanalysis}, except 
that our emphasis is on dealing with realistic foregrounds rather 
than the ability to analyze large data sets. Section 4 describes our 
plans to conduct this foreground removal challenge. 
The WOMBAT Challenge promises to shed light on several open questions 
in CMB data analysis:  What are the best foreground subtraction techniques?  
Will they allow instruments such as MAP and Planck to achieve the 
precision in $C_\ell$ reconstruction 
which has been advertised, or will the error bars increase
significantly due to uncertainties in foreground models?  Perhaps most 
importantly, do some CMB analysis methods produce biased estimates 
of the radiation power spectrum and/or cosmological parameters?  

\section{Microwave Foregrounds}	

Phase information is now available for 
Galactic dust 
and synchrotron  
and for the brightest radio galaxies, 
infrared galaxies, and X-ray clusters on the sky.  
By incorporating known information 
on the spatial distribution of the foreground components and spatial 
variation in their spectral index, we will greatly improve
upon previous highly-idealized foreground models.

There are four major expected sources of Galactic foreground emission at 
microwave frequencies:  thermal emission from dust, electric or 
magnetic dipole emission 
from spinning dust grains (Draine \& Lazarian 1998a,1998b), free-free 
emission from ionized hydrogen, and synchrotron radiation from electrons 
accelerated by the Galactic magnetic field.  Good spatial templates 
exist for thermal dust emission (Schlegel, Finkbeiner, \& Davis 1998) 
and synchrotron emission (Haslam et al. 1982), although the $0\fdg5$
resolution of the Haslam maps means that smaller-scale structure must 
be simulated.  
Extrapolation to microwave frequencies 
is possible using maps which account for spatial 
variation of the spectra (Finkbeiner, Schlegel, \& Davis 1998; 
Platania et al. 1998).  
The COMBAT collaboration has recently posted a software 
package called 
FORECAST\footnote{Foreground and CMB Anisotropy Scan Simulation Tools,\\ 
http://cfpa.berkeley.edu/group/cmbanalysis/forecast}  
 that displays the expected dust foreground for a 
given frequency, location, and observing strategy.  
Our best-fit foreground maps will be added to this user-friendly site 
in the near future, and this should be a useful resource for planning and 
simulating CMB anisotropy observations.  

A spatial template for free-free emission 
based on observations of H$\alpha$ (Smoot 1998, Marcelin et al. 1998) 
can be created in the near future 
by combining WHAM observations (Haffner, Reynolds, \& Tufte 1998) 
with the southern celestial hemisphere H-Alpha Sky Survey (McCullough 1998).   
While it is known that there is an anomalous 
component of Galactic emission at 15-40 GHz (Kogut et al. 1996, 
Leitch et al. 1997, 
de Oliveira-Costa et al. 1997) which is partially correlated 
with dust morphology, 
it is not yet clear whether this is spinning dust 
grain emission or free-free emission somehow  
uncorrelated with H$\alpha$ observations.  In fact, spinning dust grain 
emission has yet to be observed, so the uncertainties in its amplitude are 
tremendous.  A template for the ``anomalous'' 
emission component will undoubtedly have large uncertainties.  

Three nearly separate categories of galaxies will also generate microwave 
foreground emission; they are radio-bright galaxies, low-redshift 
infrared-bright galaxies, and high-redshift infrared-bright galaxies.  
The level of anisotropy produced by these 
foregrounds is predicted by Toffolatti et al. (1998) using 
models of galaxy evolution to produce source counts, and 
updated models calibrated to recent SCUBA observations are also 
available (Blain, Ivison, Smail, \& Kneib 1998, Scott \& White 1998).  
For the high-redshift galaxies detected by SCUBA, no spatial template 
is available, so a simulation of these galaxies with realistic 
clustering will be necessary.  Scott \& White (1998) 
and Toffolatti et al. (1998) 
have used very different estimates of clustering to produce divergent
results for its impact, so this issue will need to be looked at more 
carefully.  Upper and lower limits on the anisotropy generated by 
high-redshift galaxies and as-yet-undiscovered types of point sources 
are given by Gawiser, Jaffe, 
\& Silk (1998) using recent observations over a wide range of microwave 
frequencies.  Their upper limit of $\Delta T/T=10^{-5}$ for a 
$10'$ beam at 100 GHz is a sobering result; while the real sky would 
need to conspire against us to produce this much anisotropy it cannot 
be ruled out at present, and we will need to look for it with direct 
observations and design analysis techniques that might manage 
to subtract it.  
The 5319 brightest low-redshift IR galaxies detected at 60$\mu$m are 
contained 
in the IRAS 1.2 Jy catalog (Fisher et al. 1995)
 and can be extrapolated to 100 GHz with a systematic 
uncertainty of a factor of a few (Gawiser \& Smoot 1997).  
This method needs to be improved to account for the spectral difference 
between Ultraluminous Infrared Galaxies and normal spirals.  
Sokasian, 
Gawiser, \& Smoot (1998) 
have compiled a catalog of 2200 bright radio sources, 758 of 
which have been observed at 90 GHz and 309 of which have been observed 
at frequencies above 200 GHz.  They have developed a method to extrapolate
radio source spectra which has a factor of two systematic 
uncertainty at 90 GHz.
  Radio source variability represents a major challenge 
for most foreground subtraction techniques, and the information 
present in this catalog allows one to estimate the mean and variance 
of the source fluxes as a function of frequency.  

The secondary CMB anisotropies that 
occur when the photons of the Cosmic Microwave 
Background radiation are scattered after the original last-scattering 
surface can be viewed as a type of foreground contamination.
  The shape of the blackbody 
spectrum can be altered through 
inverse Compton scattering by the thermal Sunyaev-Zel'dovich (SZ) effect
 (Sunyaev \& Zel'dovich 1972).  The effective temperature of 
the blackbody can be shifted locally by 
a doppler shift from the peculiar velocity of the scattering medium (the 
kinetic SZ and Ostriker-Vishniac effects) as well as by passage through
nonlinear structure (the Rees-Sciama effect).  
Secondary anisotropies can be treated as a type of foreground contamination.
Simulations have been made of the impact of the 
SZ effects in large-scale structure (Persi et al. 1995), 
clusters (Aghanim et al. 1997), 
groups (Bond \& Myers 1996), and reionized patches
(Aghanim et al. 1996, 
Knox, Scoccimarro, \& Dodelson 1998, Gruzinov \& Hu 1998, Peebles \& 
Juskiewicz 1998).  
The brightest 200 X-ray clusters are known from the 
XBACS catalog and can be used to incorporate the locations of 
the strongest SZ sources (Refregier, Spergel, \& Herbig 1998).  
The SZ effect itself is independent of redshift, so it can yield 
information on clusters at much higher redshift than does X-ray 
emission.  However, nearly all clusters are unresolved for $10'$ resolution 
so higher-redshift clusters occupy less of the beam and therefore their SZ
effect is in fact dimmer.  In the 4.5$'$ channels of Planck this will 
no longer be true, and SZ detection and subtraction becomes  
more challenging and potentially more fruitful as a probe 
of cluster abundance at high redshift.  

\section{Reducing Foreground Contamination}

Various methods have been proposed for reducing foreground contamination.  
For point sources, it is possible to mask pixels which represent 
positive $5 \sigma$ fluctuations since such fluctuations 
are highly unlikely for Gaussian-distributed
 CMB anisotropy and 
can be assumed to be caused by point sources.    
This pixel masking technique can 
be improved somewhat by filtering (Tegmark \& 
de Oliveira-Costa 1998; see Tenorio et al. 1998 for 
a different technique using wavelets).  
Sokasian, Gawiser, \& Smoot (1998) 
demonstrate that using prior information from good source catalogs 
may allow the masking of pixels which contain sources brighter than  
the $1 \sigma$ level of CMB fluctuations and instrument noise.  For 
 the 90 GHz MAP channel, this could reduce 
the residual radio point source 
contamination by a factor of two, which 
might significantly reduce systematic errors in cosmological 
parameter estimation.  
Galactic foregrounds with well-understood frequency spectra can be 
projected out of multi-frequency observations on a pixel-by-pixel basis 
(Dodelson \& Kosowsky 1995, Brandt et al. 1994).  Prior information in 
the form of spatial templates can be included in this projection, but 
uncertainty in the spectral index is a cause for concern.  

Perhaps surprisingly, the methods for foreground subtraction which have 
the greatest level of mathematical sophistication and have been tested 
most thoroughly ignore the known locations on the sky of some  
foreground components.  The multi-frequency Wiener filtering approach 
uses assumptions about the spatial power spectra and frequency spectra of the 
foreground components to perform a separation in spherical 
harmonic or Fourier space (Tegmark \& Efstathiou 1996; 
Bouchet et al. 1995,1997,1998; Knox 1998).  However, it does not include 
any phase information at present.    
The Fourier-space Maximum Entropy Method 
(Hobson et al. 1998a) 
can add phase information on diffuse Galactic foregrounds in 
small patches of sky but treats 
extragalactic point sources as an additional source of instrument noise, 
with good results for simulated Planck data (Hobson et al. 1998b)
 and worrisome systematic 
difficulties for simulated MAP data (Jones, Hobson, \& Lasenby 1998).  
Maximum Entropy has not yet been adapted to handle full-sky 
datasets.  Both methods have difficulty if pixels are 
masked due to strong point source contamination or the spectral 
indices of the foreground components are not well known (Tegmark 1998).   

Since residual foreground contamination can increase 
uncertainties and bias parameter estimation, it is important to 
reduce it as much as possible.  Current analysis methods usually 
rely on cross-correlating the CMB maps with foreground templates at 
other frequencies (see de Oliveira-Costa et al. 1998; 
Jaffe, Finkbeiner, \& Bond 1998). 
It is clearly superior to have 
region-by-region (or pixel-by-pixel) information on how to extrapolate
these templates to the observed frequencies; otherwise this cross-correlation
only identifies the emission-weighted 
average spectral index of the foreground 
from the template frequency to the observed frequency. 

Because each foreground has a non-Gaussian spatial distribution, 
the covariance matrix of its $a_{\ell m}$ coefficients is not diagonal, 
although this has often been assumed.  When a known foreground template 
is subtracted from a CMB map, it is inevitable that the correlation 
coefficient used for this subtraction will be slightly different than 
the true value.  This expected under- or over-subtraction of each 
foreground leads to off-diagonal structure in the ``noise'' covariance matrix 
of the remaining CMB map, as opposed to the contributions of expected  
CMB anisotropies and uncorrelated instrument noise, 
both of which give diagonal contributions to the covariance matrix of 
the $a_{\ell m}$.  Thus incomplete foreground subtraction, 
like $1/f$ noise, can introduce non-diagonal correlations into the 
covariance matrix of the $a_{\ell m}$.  
These correlations complicate the likelihood analysis necessary for 
parameter estimation (Knox 1998). 
Having phase information on the 
brightness and spectral index of foreground emission 
 should reduce inaccuracies in foreground subtraction, and this motivates 
us to produce the best estimates we can of these quantities along with 
estimates of their uncertainties.  

\section{The WOMBAT Challenge}

Our purpose in conducting a ``hounds and hares'' exercise is to simulate the 
process of analyzing microwave skymaps 
as accurately as possible.  In real-world 
observations the underlying cosmological parameters 
and the exact amplitudes and spectral indices of the foregrounds are unknown, 
so Nature is the hare and cosmologists are the hounds.  
We will make our knowledge of the various foreground 
components available to the public, and each best-fit
foreground map will be accompanied 
by a map of its uncertainties and a discussion of possible systematic 
errors.  Each simulation of that foreground will be different 
from the best-fit map based upon a realization of those uncertainties.  
Very little is known about the spatial locations of high-redshift 
infrared-bright galaxies and high-redshift SZ-bright clusters, so WOMBAT 
will provide simulations of these components.  The rough characteristics 
of these high-redshift foreground sources, but not their locations, 
will be revealed.      
This simulates the real observing 
process in a way not achieved by previous foregrounds simulations.

We will release our simulated 
maps for the community to subtract the foregrounds and 
extract cosmological information. 
The WOMBAT Challenge 
is scheduled to begin on March 1, 1999 and will 
offer participating groups four 
months to analyze the skymaps and report their 
results.\footnote{see http://astro.berkeley.edu/wombat for timeline, 
details for participants, and updates}  
We will produce simulations analogous 
to high-resolution balloon observations (e.g. MAXIMA and 
BOOMERANG; see Hanany et al. 1998 and de Bernardis \& Masi 1998) 
and to the MAP 
satellite\footnote{http://map.gsfc.nasa.gov}.  
This will indicate how close the community 
is to being able to handle datasets as large as that of MAP (10$^6$ 
pixels at 13$'$ resolution for a full-sky map).  
Given current computing power, complex algorithms 
appear necessary for analyzing full-sky MAP datasets (Oh, Spergel, \& 
Hinshaw 1998), although simpler approximations may 
be possible (e.g. Wandelt, Hivon, \& G\'orski 1998).  
We plan to use 
the publicly available HEALPIX package of pixelization and analysis 
routines\footnote{http://www.tac.dk/\~{}healpix}.  
We will provide a calibration 
map of CMB anisotropy with a disclosed angular power spectrum 
in January 1999 so that participants can test the download procedure 
and become familiar with HEALPIX.  
Groups who analyze the Challenge maps 
will be asked to provide us with a summary of their analysis techniques. 
 They may choose to remain anonymous in our comparison of the results but  
are encouraged to publish their own conclusions  
based on their participation.  

One of the biggest challenges in real-world observations is being prepared 
for surprises, both instrumental and astrophysical (see 
Scott 1998 for an eloquent discussion).  An exercise 
such as the WOMBAT Challenge 
is an excellent way to simulate these surprises, and we will 
include a few in our skymaps.  
The results of the WOMBAT Challenge will provide estimates of the 
 effectiveness of  current 
techniques of foreground subtraction, power spectrum analysis, and 
parameter estimation.  

\section{Conclusions}

Undoubtedly the most important scientific contribution that WOMBAT will 
make is the production of realistic full-sky maps of all major microwave 
foreground components with estimated uncertainties.  These maps are needed 
for foreground subtraction and estimation of residual foreground contamination 
in present and future CMB anisotropy observations.  They will allow 
instrumental teams to conduct realistic simulations of the observing and 
data analysis process without needing to assume overly idealized models 
for the foregrounds.  By combining various realizations of these 
foreground maps within the stated uncertainties with a simulation of 
the intrinsic CMB anisotropies, we will produce the best simulations 
so far of the microwave sky.  Using these simulations in a ``hounds and 
hares'' exercise should test how 
well the various foreground subtraction and parameter estimation techniques
work at present.  It is easy 
to question 
the existing tests of analysis methods which assume idealized foregrounds 
in analyzing similarly idealized simulations.

Data analysis techniques will undoubtedly improve with time, and we hope 
to reduce the current uncertainty in their efficacy 
such that follow-up simulations 
by the instrumental teams themselves can generate confidence in 
the results of real observations.  
We can test the resilience of CMB analysis methods to surprises such as 
unexpected foreground 
amplitude or spectral behavior, correlated instrument noise, and 
CMB fluctuations from non-gaussian or non-inflationary models.  
Cosmologists need to know if 
such surprises can lead to the misinterpretation of cosmological 
parameters.  
  In the future, we envision producing 
time-ordered data, simulating interferometer observations, and adding 
polarization to our microwave sky simulations.  

Perhaps the greatest advance we offer is the ability to evaluate the 
importance of studying the detailed locations of foreground sources.  
If techniques which ignore this phase information are still successful 
on our realistic sky maps, that is a significant vote of confidence.  
Alternatively, it may turn out that techniques which use 
phase information are needed in order to reduce foreground contamination 
to a level which does not seriously bias the estimation of cosmological 
parameters.  Combining various techniques may lead to improved 
foreground subtraction methods, and we hope that a wide variety of 
techniques will be tested by the participants in the WOMBAT Challenge.

\section{Acknowledgments}

We thank Rob Crittenden (IGLOO) and Kris Gorski, Eric Hivon,
and Ben Wandelt (HEALPIX) for making pixelization schemes available 
to the community.  We appreciate helpful conversations with Nabila 
Aghanim, Giancarlo de Gasperis, Mark Krumholz, Alex Refregier, and Philip 
Stark and gratefully acknowledge the support of NASA LTSA grant \#NAG5-6552.


\begin{references}

\reference{Aghanim et al. 1997}
Aghanim, N., De Luca, A., Bouchet, F. R., Gispert, G., \& Puget, J. L. 1997, 
\aap, 325, 9, astro-ph/9705092

\reference{Aghanim et al. 1996}
Aghanim, N., Desert, F. X., Puget, J. L., Gispert, R. 1996, \aap, 311, 1





\reference{Blain, Ivison, Smail, \& Kneib 1998}
Blain, A. W., Ivison, R. J., Smail, I., \& Kneib, J. P. 1998, 
 to appear in {\it Wide-field surveys in cosmology, Proc. XIV IAP meeting}, 
 astro-ph/9806063 

\reference{Bond, Efstathiou, \& Tegmark 1997}
Bond, J. R., Efstathiou, G., \& Tegmark, M. 1997, \mnras 291, L33

\reference{Bond \& Myers 1996}
Bond, J. R., \& Myers, S. T. 1996, \apjs, 103, 63

\reference{Bouchet et al. 1997}
Bouchet, F. R., Gispert, R., Boulanger, F., \& Puget, J. L. 1997, 
in Bouchet F. R., Gispert R., Guiderdoni, B., Tran Thanh Van J., eds., 
Proc. 36th Moriond Astrophysics Meeting, Microwave Anisotropies, 
Editions Frontiere, Gif-sur-Yvette, p. 481

\reference{Bouchet et al. 1995}
Bouchet, F. R., Gispert, R. \& Puget, J. L. 1995, 
in ``Unveiling the Cosmic Infrared Background,'' AIP Conference Proceedings
348, Baltimore, MD, USA, ed. E. Dwek, p.255

\reference{Bouchet et al. 1998}
Bouchet, F. R., Prunet, S., \& Sethi, S. K. 1998, astro-ph/9809353

\reference{Brandt et al. 1994}
Brandt, W. N., Lawrence, C. R., Readhead, A. C. S., Pakianathan, J. N.,
\& Fiola, T. M. 1994, \apj, 424, 1 




\reference{de Bernardis \& Masi 1998}
de Bernardis, P., Masi, S. 1998, to appear in ``Fundamental 
parameters in Cosmology,'' Rencontres de Moriond,
astro-ph/9804138

\reference{de Oliveira-Costa et al. 1997}
de Oliveira-Costa, A., et al. 1997, \apj, 482, L17

\reference{de Oliveira-Costa et al. 1998}
de Oliveira-Costa, A., Tegmark, M., Page, L. A., \& Boughn, S. P. 
1998, \apjl, in press, astro-ph/9807329

\reference{Dodelson \& Kosowsky 1995}
Dodelson, S. \& Kosowsky, A. 1995, \prl, 75, 604

\reference{Draine \& Lazarian 1998a}
Draine, B. T. \& Lazarian, A., 1998a, \apj, 494, L19

\reference{Draine \& Lazarian 1998b}
Draine, B. T. \& Lazarian, A., 1998b, astro-ph/9807009


\reference{Eisenstein, Hu, \& Tegmark 1998}
Eisenstein, D. J., Hu, W., \& Tegmark, M. 1998, astro-ph/9807130


\reference{Finkbeiner et al. 1998}
Finkbeiner, D., Schlegel, D., Davis, M. 1998, in preparation 

\reference{Fisher et al. 1995}
Fisher, K. B., Huchra, J. P., Strauss, M. A., Davis, M., Yahil, A., 
\& Schlegel, D. 1995, \apjs, 100, 69 




\reference{GJS}
 Gawiser, E., Jaffe, A., \& Silk, J.  1998, astro-ph/9811148

\reference{GS97}
 Gawiser, E. \& Smoot, G. F. 1997, \apjl, 480, L1 


\reference{Gruzinov \& Hu 1998}
Gruzinov, A., \& Hu, W. 1998, astro-ph/9803188

\reference{Haffner, Reynolds, \& Tufte 1998}
Haffner, L. M., Reynolds, R. J., \& Tufte, S. L. 1998, \apj, 501, L83

\reference{Hanany et al. 1998}
Hanany, S. et al. 1998, 
Proc. of 18th Texas Symposium on Relativistic Astrophysics and 
Cosmology, ed. A. V. Olinto, J. A. Frieman, \& D. N. Schramm, 
World Scientific, p.255

\reference{Haslam et al. 1982}
Haslam, C. G. T., Salter, C. J., Stoffel, H., \& Wilson, W. E. 1982, \aaps, 
47, 1 

\reference{Hobson et al. 1998a}
Hobson, M. P., Jones, A. W., Lasenby, A. N., \& Bouchet, F. R.  1998a, 
\mnras, in press

\reference{Hobson et al. 1998b}
Hobson, M. P., Barreiro, R. B., Toffolatti, L., Lasenby, A. N., Sanz, 
J. L., Jones, A. W., \& Bouchet, F. R. 1998b, astro-ph/9810241





\reference{Jaffe, Finkbeiner, \& Bond 1998}
Jaffe, A., Finkbeiner, D., \& Bond, J. R. 1998, in preparation

\reference{Jones, Hobson, \& Lasenby 1998}
Jones, A. W., Hobson, M. P., \& Lasenby, A. N. 1998, astro-ph/9810236

\reference{Jungman et al. 1996}
Jungman, G., Kamionkowski, M., Kosowsky, A., \& Spergel, D. N. 1996,
 \prd, 54, 1332

\reference{Knox 1998}
Knox, L. 1998, astro-ph/9811358

\reference{Knox, Scoccimarro, \& Dodelson 1998}
Knox, L., Scoccimarro, R., \& Dodelson, S. 1998, \prl, 81, 2004, 
astro-ph/9805012

\reference{Kogut et al. 1996}
 Kogut, A. et al. 1996, \apjl, 464, L5

\reference{Leitch et al. 1997}
Leitch, E. M., Readhead, A. C. S., Pearson, T. J., Myers, S. T. 1997, 
\apj, 486, L23, astro-ph/9705241



\reference{Marcelin et al. 1998}
Marcelin, M., Amram, P., Bartlett, J. G., Valls-Gabaud, D., \& 
Blanchard, A. 1998, \aap, 338, 1 

\reference{McCullough 1998}
McCullough, P. 1998, to appear in Proceedings of Sloan Summit on 
Microwave Foregrounds, in preparation

\reference{Oh, Spergel, \& Hinshaw 1998}
Oh, S. P., Spergel, D. N., \& Hinshaw, G. 
1998, \apj, in press, astro-ph/9805339

\reference{Peebles \& Juskiewicz 1998}
Peebles, P. J. E., \& Juskiewicz, R. 1998, astro-ph/9804260

\reference{Persi et al. 1995}
Persi, F. M., Spergel, D. N., Cen, R., \& Ostriker, J. P. 1995, \apj, 
442, 1 


\reference{Platania et al. 1997}
Platania, P. et al., 1997, \apj, 505, 473 



\reference{Refregier et al. 1998}
 Refregier, A., Spergel, D. N, \& Herbig, T., 1998, astro-ph/9806349




\reference{Schlegel, Finkbeiner, Davis 1998}
Schlegel, D., Finkbeiner, D., \& Davis, M. 1998, \apj, 500, 525

\reference{Scott 1998}
Scott, D., 1998, to appear in Proc of the 
MPA/ESO Conference:  ``Evolution of Large-Scale Structure:  from 
Recombination to Garching,'' ed. A. J. Banday et al, 1998, 
astro-ph/9810330

\reference{Scott \& White 1998}
Scott, D., \& White, M. 1998, astro-ph/9808003



\reference{Smoot 1998}
 Smoot, G. F. 1998, astro-ph/9801121

\reference{SGMS}
 Sokasian, A., Gawiser, E., \& Smoot, G.F. 1998, astro-ph/9811311


\reference{SZ}
 Sunyaev, R.A. \& Zel'dovich, Ya.B. 1972, {\it Comments Ap. Space Sci.} 4, 173 

\reference{Tegmark 1998}
 Tegmark, M. 1998, \apj, 502, 1 

\reference{Tegmark \& de Oliveira-Costa 1998}
 Tegmark, M. \& de Oliveira-Costa, A. 1998, \apj, 500, L83

\reference{Tegmark \& Efstathiou, 1996}
 Tegmark, M. \& Efstathiou, G. 1996, \mnras, 281, 1297

\reference{Tenorio et al. 1998}
Tenorio, L., Lineweaver, C., Hanany, S., \& Jaffe, A., 1998, in preparation  


\reference{Toffolatti et al. 1998}
 Toffolatti, L. et al. 1998, \mnras, 297, 117

\reference{Wandelt, Hivon, \& Gorski 1998}
Wandelt, B. D., Hivon, E., \& G\'orski, K. M. 1998, astro-ph/9808292



\reference{Zaldarriaga, Spergel, Seljak 1997}
Zaldarriaga, M., Spergel, D. N., \& Seljak, U. 1997, astro-ph/9702157


\end{references}
\end{document}